\documentclass[global,twocolumn,final]{svjour}
\usepackage{latexsym}
\usepackage{revsymb}
\usepackage{amssymb}
\usepackage{graphics}
\journalname{Applied Physics B}
\begin{document}
\title{Nanoscale atomic waveguides with suspended carbon nanotubes}
\author{V. Peano \inst{1} \and M. Thorwart\inst{1} \and A.\ Kasper\inst{2}
 \and R. Egger\inst{1}}
%
%
\institute{Institut f\"ur Theoretische Physik,
Heinrich-Heine-Universit\"at D\"usseldorf, D-40225 D\"usseldorf,
Germany \and NTT Basic Research Laboratories, NTT Corporation, Kanagawa 243-0198,
Japan}
\date{Received:  / Revised version: }
%
\maketitle
\begin{abstract}
We propose an experimentally viable setup for the realization of
one-dimensional ultracold atom ga\-ses in a nanoscale magnetic
waveguide formed by single dou\-bly-clamped suspended carbon
nano\-tubes. We show that all common decoherence and atom loss
mechanisms are small guaranteeing a stable operation of the trap. 
Since the extremely large current densities in carbon nanotubes are
spatially homogeneous, 
our proposed architecture allows to 
create a very regular trapping potential for 
the atom cloud. 
Adding a second nanowire allows to create a double-well 
potential with a moderate tunneling barrier which is desired for
tunneling and interference experiments with the advantage of
tunneling distances being in the nanometer regime. 
\end{abstract}

\section{Introduction}
\label{intro} The ongoing progress in the fabrication and
manipulation of micro- or nanoscale structures has recently allowed
for systematic studies of ultracold atom gases, where
current-carrying wires and additional magnetic bias fields 
generate magnetic fields trapping neutral
atoms (`atom chips')  \cite{Folman02,Reichel02}. 
For instance, the Bose-Einstein condensation (BEC) of
microchip-con\-fined atoms has been successfully demonstrated by
several groups \cite{atomBEC}.  
So far, deco\-her\-ence and atom loss constitute central
im\-pe\-di\-ments, since atoms are relatively close to `hot'
macroscopic surfaces or current-carrying wires (with typical
diameters of several $\mu$m), where the Casimir-Polder potential and
Johnson noise can seriously affect stability
\cite{Henkel99,chin,Schroll03}. To reduce these effects, further
miniaturization to the nanoscale regime would be desirable. In
particular, this is 
promising in the context of integrated atomic matter-wave
interferometry and optics \cite{Kasevich}, and combines the
strengths of nanotechnology and atomic physics. 
While at first sight this goal conflicts with the requirement of
large currents forming tight trapping potentials, 
we propose that when
using suspended carbon nanotubes (NTs) \cite{tubes} (with diameters
of a few nm) as  wires, nanoscale atom chip devices with large current
densities can be designed. In turn, these devices allow to trap
ultracold atom gases basically free
of trap-induced decoherence or atom losses, with the gas containing
few tens of atoms. Since disorder is generally weak in NTs, the (extremely
large) current density distribution is spatially homogeneous, which allows
to overcome the problem of fragmentation of the atom cloud. 
Moreover, they can be built with state-of-the-art
technology. 

With relevant length scales below optical and cold-atom de Broglie
wavelengths, this also paves the way for the observation of
interesting and largely unexplored many-body physics in one
dimension (1D) \cite{Petrov04}. Examples include the interference
properties of interacting matter waves \cite{chen}, the 1D analogue
of the BEC-BCS crossover \cite{becbcs} and shape resonances in 1D
trapping potentials \cite{olshanii}. Previous realizations of 1D
cold atoms were reported using optical lattices
\cite{esslinger,Paredes04,Weiss} and magnetic traps
\cite{Goerlitz01}, but they involve arrays of 1D or elongated 3D
systems, where it is difficult to separately 
manipulate  a single 1D atom cloud  (the distances between the 1D
 systems composing the array 
are few hundred  nm).  A noteworthy advantage of our proposal  against
dipole optical traps  is that
arrays of many NT waveguides can be built, where it is possible to
manipulate an individual trap by changing the current through an 
individual NT. Moreover,  our proposal allows to  
 minimize unwanted substrate effects and implies a
drastically reduced transverse size (a few nm) of the cloud. 
We expect that our approach allows to observe new interesting many-body
features not accessible by
the otherwise very successful atom chip setup. This could provide a 
fruitful link between atomic and condensed matter physics 
with a wealth of fascinating effects. 

\section{The setup}

\begin{figure}[h!]
\begin{center}
\resizebox{0.37\textwidth}{!}{%
  \includegraphics{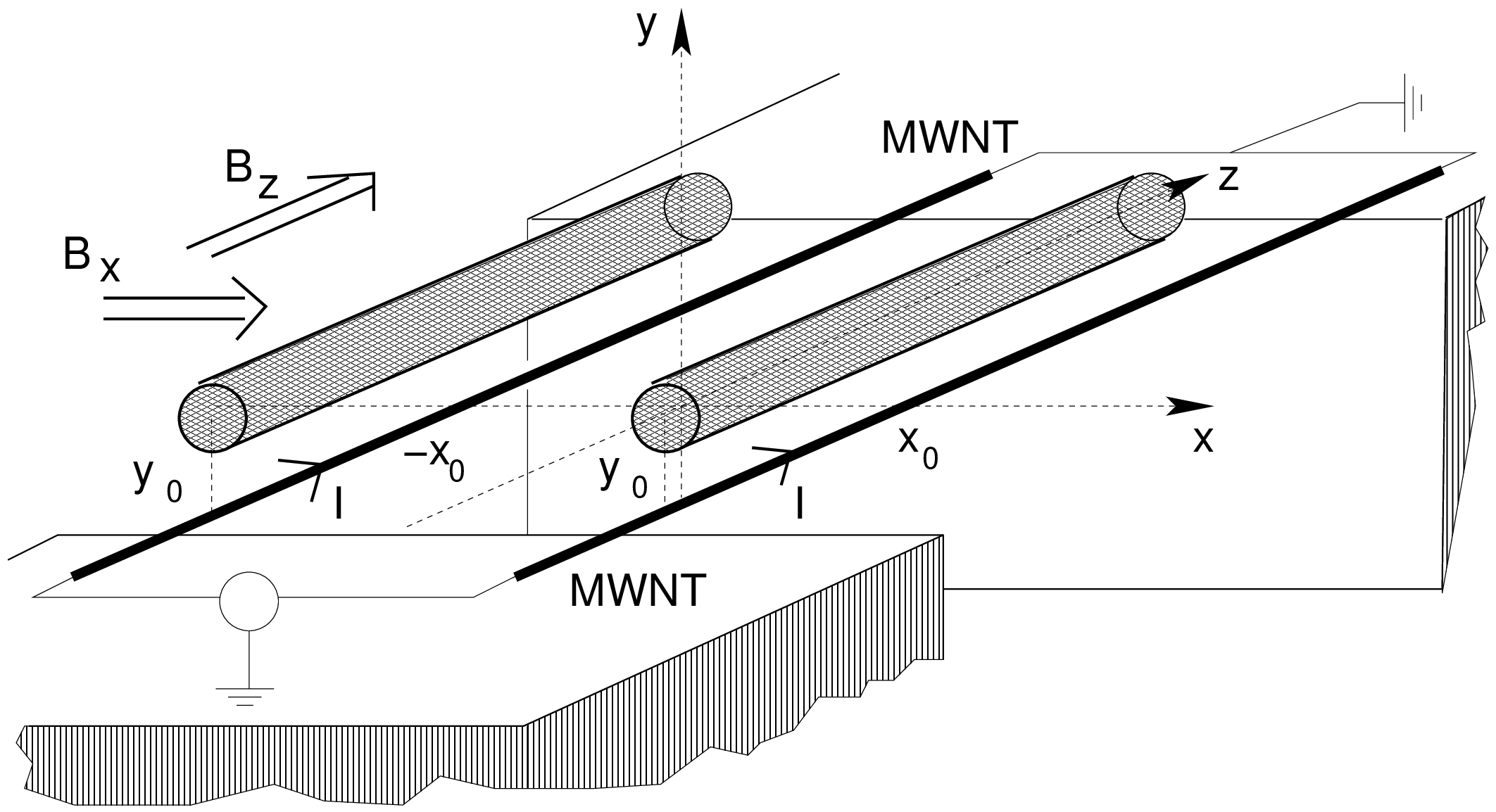}
}
\caption{Sketch of the proposed device. A current-carrying
suspended NT is positioned at $(-x_0,0,z)$ and together with the
transverse magnetic field $B_x$, a 1D trapping potential is formed.
The shaded region indicates the atom gas. A similar two-wire setup
allows the creation of a bistable potential. \label{fig.1}}
\end{center}
\end{figure}

A typical proposed nanoscale waveguide setup to confine ultracold
atoms to 1D is sketched in Fig.~\ref{fig.1}. The setup employs a
single suspended doubly-clamped NT (left NT in Fig.~\ref{fig.1}, the
second suspended NT on the right will be used to create a
double-well potential, see below), where nanofabrication techniques
routinely allow for trenches with typical depths and lengths of
several $\mu$m \cite{tubes}. To minimize decoherence and loss
effects \cite{chin}, the substrate should be insulating apart from
thin metal strips to electrically contact the NTs. Since strong
currents (hundreds of $\mu$A)  are necessary, thick
 multiwall nanotubes (MWNTs) or `ropes'  \cite{tubes} are best suited.
The suspended geometry largely eliminates the influence of the
substrate. A transverse magnetic field $B_x$ is required to create a
stable trap while a longitudinal magnetic field $B_z$
suppresses Majorana spin flips \cite{sukumar,jones}. 
With this single-tube setup, neutral atoms in a weak-field seeking
state can be trapped. Studying various sources for decoherence,
heating or atom loss, and estimating the related time scales, we
find that, for reasonable parameters, detrimental effects are small.
As a concrete example, we shall consider $^{87}$Rb atoms in the
weak-field seeking hyperfine state $|F,m_F\rangle=|2,2\rangle$.

We next describe the setup in Fig.~\ref{fig.1}, where the
(homogeneous) current $I$ flows through the left NT positioned at
$(-x_0,0,z)$. With regard to the decoherence properties of the proposed trap,
it is advantageous  that the current flows 
homogeneously through the NT, as disorder effects are usually weak
in NTs \cite{tubes}.  
Neglecting boundary effects due to the finite tube length $L$, the
magnetic field at  ${\bf x}=(x,y,z)=({\bf x}_\perp,z)$ is given by
\begin{equation}
\mathbf{B}({\bf x})=\frac{\mu_0 I}{2\pi}
\frac{1}{(x+x_0)^2+y^2} \left( \begin{array}{c} -y \\ x+x_0 \\ 0\\
\end{array}\right) +\left( \begin{array}{c} B_x\\0\\B_z\\\end{array}\right)
\end{equation}
with the vacuum permeability $\mu_0$. To create a trapping potential
minimum at $y_0$, the transverse magnetic field is 
$B_x=\mu_0 I / (2\pi y_0)$. Then
the transverse confinement potential is
$V({\bf x}_\perp)=\mu |{\bf B({\bf x})}|$, where $\mu=m_F g_F \mu_B$
 with the Land{\'e} factor $g_F$ and the Bohr magneton $\mu_B$.
 It has a minimum along the line $(-x_0,y_0,z)$, with the distance between
 the atom cloud and the wire being $y_0$.
Under the adiabatic approximation \cite{sukumar},
$m_F$ is a constant of motion, and the
potential is harmonic very close to the minimum of the trap, i.e.,
$V(\mathbf{x})\simeq \mu B_z+\frac{1}{2}m\omega^2
[(x+x_0)^2+(y-y_0)^2]$, with
frequency $\omega=  [\mu/(m B_z)]^{1/2} \mu_0 I / (2\pi y_0^2)$
and associated transverse confinement length $l_0=(\hbar/m\omega)^{1/2}
\ll y_0$, where $m$ is the atom mass.
The adiabatic approximation is valid as long as $\omega\ll \omega_L$
 with the Larmor frequency $\omega_L= \mu B_z/\hbar$.
Non-adiabatic
Majorana spin flips to a strong-field seeking state generate atom loss 
\cite{Folman02,jones}  characterized by the rate  $\Gamma_{\rm loss} \simeq
(\pi\omega/2)\exp(1-1/\chi)$, with $\chi=\hbar\omega/(\mu B_z)$
\cite{sukumar}. For convenience, we switch to a dimensionless form
of the full potential $V({\bf x}_\perp)$ by measuring energies in
units of $\hbar \omega$ and lengths in units of $l_0$,
\begin{equation} \label{fullpot}
\chi V= \left( 1+\chi \frac { d^2 [ (x+x_0)^2-dy+y^2]^2
+ d^4(x+x_0)^2 }
{ [(x+x_0)^2+y^2]^2 } \right)^{1/2},
\end{equation}
which depends only on $d= y_0/l_0$ and $\chi$. The trap
frequency then follows as
\begin{equation}
\omega=\frac{m\chi\mu^2}{\hbar^3}\Big(\frac{\mu_0I}{2\pi d^2}\Big)^2
\, .
\end{equation}
Note that a real trap also requires a longitudinal
confining potential with frequency $\omega_z \ll \omega$. 

\begin{figure}[h!]
\begin{center}
a1)
\resizebox{0.20\textwidth}{!}{%
  \includegraphics{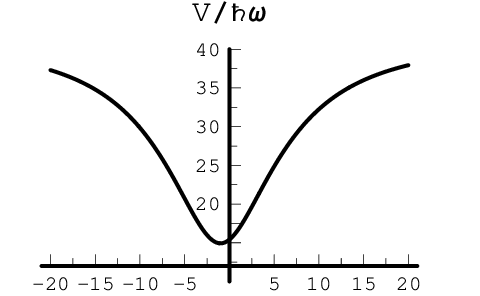}
}
\hfill
b1)
\resizebox{0.20\textwidth}{!}{%
  \includegraphics{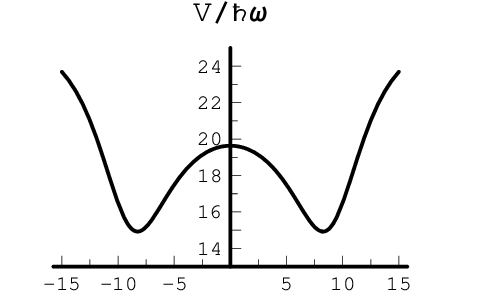}
}\\
a2)\hspace*{-0.3cm}
\resizebox{0.20\textwidth}{!}{%
  \includegraphics{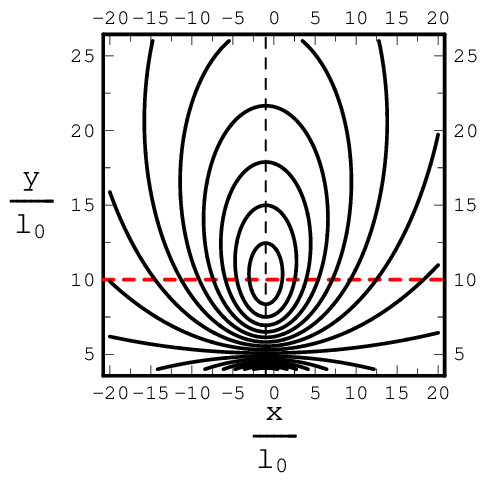}
}
\hfill
b2)\hspace*{-0.3cm}
\resizebox{0.20\textwidth}{!}{%
  \includegraphics{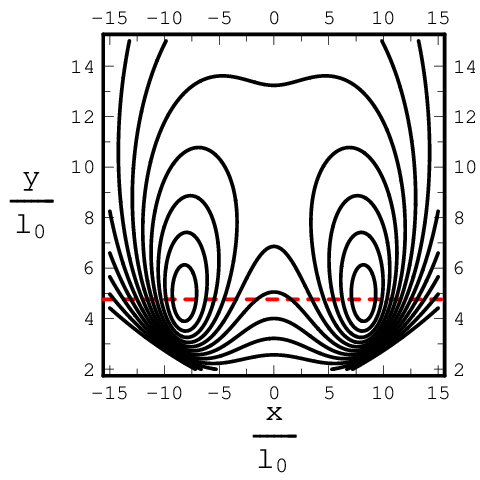}
}\hspace*{0.3cm}
\caption{(a) Transverse trapping potential of the nanoscale
waveguide for $I=100$ $\mu$A, $d=10$, $\chi=0.067$ and $x_0=l_0$. The
resulting trap frequency is $\omega = 2 \pi \times 4.6$ kHz while 
$y_0=1440$ nm, corresponding to $B_x=0.14$ G.  (a1) shows a cut along $y=y_{\rm min}$ 
through the contour plot shown in (a2), see horizontal dashed line.  
(b) Bistable potential for the double-wire configuration for
$\chi=0.067, I=100$ $\mu$A, $x_0=200$ nm and $y_0=100$ nm. (b1) displays a cut 
along $y=y_{\rm min}$  through the contour shown in (b2), see dashed line. 
\label{fig.2}}
\end{center}
\end{figure}

To obtain an estimate for the design of the nanotrap, we
 choose realistic parameters: $\chi=0.067$, corresponding to a rate of
spin flip transition per oscillation period
$\Gamma_{\rm loss}/\omega\sim10^{-6}$. 
Decreasing $d$ increases the trap frequency. However, $d$ cannot be
chosen too small, for otherwise the potential is not confining
anymore (and the harmonic approximation becomes invalid). 
Using $V(\infty)=\chi^{-1}(1+\chi
d^2)^{1/2}$ for the potential at $|{\bf x}_\perp|\to \infty$, we now show
that for $d\alt 5$, the harmonic approximation
breaks down. To see this, note that for $d=10$,
 the  potential provides a confining barrier
(in units of the trap frequency $\omega$) of
$V(\infty)-V(0,0,z)=23.8$, while for $d=5$, we get only
$V(\infty)-V(0,0,z)=9.8$. Thus exceedingly
 small values of $d$ would lead to unwanted
thermal atom escape processes out of the trap.  
To illustrate the feasibility of the proposed trap design, we show
in Table \ref{tab.freq} several parameter combinations with
realistic values for the MWNT current together with the resulting
trap parameters.  In practice, first the maximum possible current
should be applied to the NT, with some initial field $B_x$.  After
loading of the trap, the field $B_x$ should be increased, the cloud
thereby approaching the wire with a steepening of the confinement.
At the same time, $y_0$ and consequently  $d$ decrease. 
This procedure can be used to load the nanotrap from a larger magnetic
trap (ensuring mode matching). 
For a given current, there is a corresponding lower limit
$y_{\rm min}$ for stable values of $y_0$ from the
requirement $d\agt 5$, as already mentioned above. 
To give an example, the confining potential is shown in
Fig.~\ref{fig.2}a) for $I=100 \mu$A, representing a reasonable
current through thick NTs \cite{tubes}, $d=10$, $x_0= l_0$ and
$\Gamma_{\rm loss}/\omega = 10^{-6}$  (where $\chi=0.067$). The
resulting trap frequency is $\omega= 2 \pi \times 4.6$ kHz and the associated
transverse magnetic field is $B_x=0.14$ G. 

\section{Influence of destructive effects}

For stable operation, it is essential that destructive effects like
atom loss, heating or decoherence are small. 

(i) One loss process
is generated by non-adiabatic Majorana spin flips as discussed
above. 

(ii) Atom loss may also originate from noise-induced spin flips,
where current fluctuations cause a fluctuating magnetic field
generating the Majorana spin flip rate \cite{Henkel99}
\begin{equation}
\gamma_{\rm sf} \simeq  \left(\frac{\mu_0\mu}{2\pi\hbar
 y_0}\right)^2 \frac{S_I(\omega_L)}{2}, \
S_I(\omega)=\int dt e^{-i\omega t}\langle I(t)I(0)\rangle.
\end{equation}
At room temperature and for typical voltages $V_0\approx 1$~V, we have
$\hbar\omega_L\ll k_BT\ll eV_0$,  and $S_I(\omega_L)$ is
expected to equal the shot noise $2eI/3$ of a diffusive wire.
For the parameters above,  a rather small escape rate results,
$\gamma_{\rm sf}\approx 0.051$~Hz. If a (proximity-induced)
supercurrent is applied to the MWNT, the resulting current fluctuations
could be reduced even further.   

(iii) Thermal NT vibrations might create decoherence  and heating,
and could even cause a transition to the first excited state of the
trap.  Using a standard elasticity model for a doubly clamped wire
in the limit of small deflections, 
the maximum mean square displacement is \cite{Sapmaz03} 
\[
\sigma^2=\langle \phi^2(L/2)\rangle=
k_B T L^3/(192 Y M_I),
\]
where $\phi(z)$ is the NT displacement, $L$ the (suspended) 
NT length, $T$ the temperature, $Y$ the Young modulus, and $M_I$
the NT's moment of inertia. For $L=10 \mu$m and typical
material parameters from Ref.~\cite{tubes}, we find $\sigma\approx
0.2$~nm at room temperature. This is much smaller than the
transverse size $l_0$ of the atomic cloud. Small
fluctuations of the trap center could cause transitions to excited
transverse trap states. Detailed analysis shows
that the related decoherence rate is also negligible, since the
transverse fundamental vibration mode of the NT has the frequency 
\begin{equation}
\omega_{\rm f} =\frac{\beta_1^2}{L^2}\sqrt{\frac{Y M_I}{\rho_L A_c}} \, , 
\end{equation}
with $\beta_1\simeq 4.73$, the mass density $\rho_L$, and the 
cross-sectional area $A_c$. For the above parameters, 
$\omega_{\rm f} = 2 \pi \times 11.9$ MHz is much larger than the trap
frequency itself. Due to the strong frequency mismatch, the coupling
of the atom gas to the NT vibrations is therefore negligible. 

(iv) Another decoherence mechanism comes from current fluctuations
 in the NTs. Following the analysis of Ref.~\cite{Schroll03}, the
corresponding decoherence rate is
\begin{equation}
\frac{\gamma_c}{\omega} = \frac{3 \pi}{4\hbar} k_B T \frac{\sigma_0
A}{y_0^3}
\left(\frac{\mu_0 \mu_B}{2 \pi}\right)^2 \frac{\chi}{\hbar \omega} \,
,
\end{equation}
where $\sigma_0$ is the NT conductivity and $A$ the cross-sectional
area through which the current runs in the NT.
For the corresponding parameters we find $\gamma_{\rm c}/\omega < 10^{-8}$.

(v) Another potential source of atom loss could be the attractive
 Casimir-Polder  force between the atoms and the NT surface.
 The Casimir-Polder interaction potential
 between an infinite plane and a neutral atom
is given by $V_{\rm CP}=-C_4/r^4$ \cite{chin,Casimir48}.
For a metallic surface and $^{87}$Rb atoms,
$C_4=1.8\times10^{-55}$ Jm$^4$, implying that at a distance of
$1\mu$m from the surface,  the characteristic frequency associated
with the Casimir-Polder interaction is $V_{\rm CP}/\hbar= 2 \pi
\times 0.29$ kHz.
In our setup, however, we cannot apply this estimate since the assumption of an 
infinite plane is not realistic for a NT with a diameter of a few tens of 
nm. Instead, 
we expect that the distance between the cloud and the NT can be reduced without 
 drastically increasing the Casimir-Polder force. 
The proposed setup could be an interesting 
playground to study the Casimir-Polder interaction for our more 
complicated geometry. 

(vi) A further possible mechanism modifying the shape of the
confining potential is the influence of the electric field 
between the two contacts of the nanowire and the macroscopic leads
which is created by the transport voltage. This field depends
strongly  on the detailed geometry of the contacts. However, the
electric field can in general be
 reduced if the total length $L_{\rm tot}$
of the NT is increased. (Note that $L_{\rm tot}$
can be different from the length $L$ over which the NT is
suspended). Due to the small intrinsic NT resistivity, the
influence of the contact resistance then decreases for longer NTs. 
Finally, we mention that superconducting leads could be used 
 to reduce the voltage drop.

\section{Number of trapped atoms and size of atom cloud}

Next we address the important issue of how many atoms
can be loaded into such a nanotrap.
This question strongly depends on
the underlying many-body physics which determines for instance the
density profile of the atom cloud. Since the trap frequencies given in Table
 \ref{tab.freq} exceed  typical thermal energies of the cloud,
we will consider the 1D situation.  Within the framework of
two-particle s-wave scattering in a parabolic trap,
the effective 1D interaction strength $g_{\rm 1D}=-2\hbar^2/ (ma_{\rm
1D})$ is related to the
3D scattering length $a$ according to \cite{olshanii}
\begin{equation}
a_{\rm 1D}=-\frac{l_0^2}{a} \left(1-{\cal
C}\frac{a}{\sqrt{2}l_0}\right) \, ,
\end{equation}
where ${\cal C}\simeq 1.4603.$ Interestingly,
$g_{\rm 1D}$ shows a confinement-induced resonance (CIR) for
 $a=\sqrt{2} l_0 / {\cal C}$ \cite{olshanii}.
For nearly parabolic traps respecting parity symmetry,
this CIR is split into three resonances  \cite{Peano}.
However,  for the typical trap frequencies displayed in
Table \ref{tab.freq}, corresponding to non-resonant
atom-atom scattering, the parabolic confinement 
represents a very good approximation. For free bosons in 1D,
the full many-body problem can be solved analytically \cite{Lieb}. It
turns out that the governing  parameter is given by
$n|a_{\rm 1D}|$, where $n$ is the atom density in the
cloud. For weak interactions (large $n|a_{\rm 1D}|$),
a Thomas-Fermi (TF) gas results, while in the opposite regime, the
Tonks-Girardeau (TG) gas is obtained. 

For realistic traps with an
additional longitudinal confining potential with frequency
$\omega_z \ll \omega$, the problem has
been addressed in Ref.\  \cite{Dunjko}. 
The corresponding governing parameter is $\eta=n_{\rm TF}|a_{1D}|$
 where $n_{\rm TF}=[(9/64)N^2(m\omega_z\hbar)^2|a_{1D}|]^{1/3}$ is the cloud
 density in the center of the trap in the TF approximation.
 Small $\eta$ characterizes
 a TG gas whereas large $\eta$ corresponds to the TF gas.
The longitudinal size $\ell$ of the atom cloud in terms of the atom
 number $N$ and the longitudinal (transversal) trap frequencies
 $\omega_z$ ($\omega$) has been computed in Ref.\
  \cite{Dunjko}, with the result
 \begin{equation}
 \ell = \left[  \frac{3N (\hbar/m\omega_z)^2}{|a_{\rm 1D}|}
\right]^{1/3}
 \label{length1}
 \end{equation}
 in the TF regime and
 \begin{equation}
 \ell = \left[
 2N (\hbar/m\omega_z)
 \right]^{1/2}
 \label{length2}
 \end{equation}
 in the TG regime.
 In order to determine the cloud size $\ell$, we first calculate
  $\eta$ for fixed  $N, \omega_z$ and $\omega$, and then use the
  respective formula Eq.\  (\ref{length1}) or
  (\ref{length2}). In the crossover region, both expressions yield
  similar results that also match the full numerical solution
  \cite{Dunjko}.
 Typical results for  realistic parameters are listed in Table
  \ref{tab.number} for $\omega_z=2\pi \times 0.1$ kHz.
{}From these results, we conclude that the length of the suspended
NT should be in the $\mu$m-regime in order to trap a few tens of
$^{87}$Rb atoms. 

To summarize the discussion of the monostable trap,
we emphasize that 
the proposed nanotrap is realistic, with currents of a few 100 $\mu$A
and lengths of few $\mu$m of the suspended parts of NT.
No serious fundamental decoherence, heating or loss mechanisms are expected for
reasonable parameters of this nanotrap. We note that we did not consider 
additional specific noise sources from further experimental equipment.  

\section{Double-well potential with two carbon nanotubes}

In order to illustrate the advantages of the miniaturization to the
nanoscale, let us consider a setup which allows two stable minima
separated by a tunneling barrier. The simplest setup consists of two
parallel NTs carrying co-propagating currents $I$, a (small)
longitudinal bias field $B_z$ and  a transverse bias field $B_x$.
Such a double-well potential for 1D ultracold atom gases would
permit a rich variety of possible applications. Experiments to study
Macroscopic Quantum Tunneling and Macroscopic Quantum
Coherence phenomena \cite{weiss}
between strongly correlated 1D quantum
gases could then be performed.
In addition, qubits forming the building blocks for a quantum
information processor could be realized. 
The rich tunability of the potential shape, including
tuning the height of the potential barrier as well as the tunneling
distance, is a particularly promising feature.

\begin{figure}[h!]
\begin{center}
\resizebox{0.38\textwidth}{!}{%
  \includegraphics{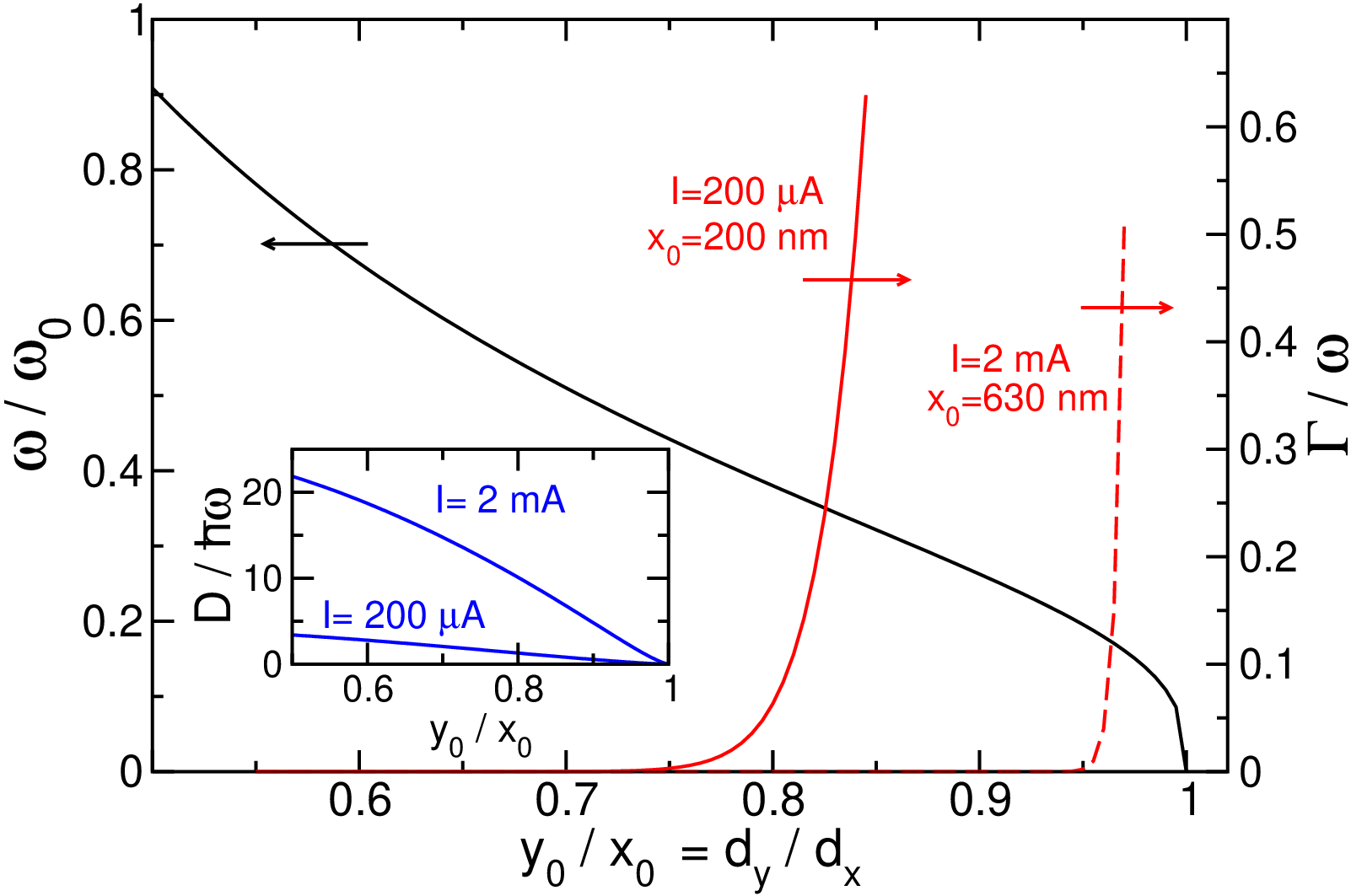}
}
\caption{(Color online). Trap frequency $\omega$ in the bistable potential (left
scale) and tunneling rate $\Gamma$ within the WKB-approximation (right
scale) as a function of the ratio $y_0/x_0=d_y/d_x$. For the
definition of $\omega_0$, see text. The tunneling rate $\Gamma$ is computed
for $^{87}$Rb
atoms with $\chi=0.067$ and $x_0=200$ nm for two values of the current
$I$ given in the figure.
 \label{fig.3}}
\end{center}
\end{figure}

To realize this potential, we propose to place  a second
current-carrying NT  at $(+x_0,0,z)$, where the condition
$x_0>y_0$ guarantees the existence of two minima
located at $y_0(\pm \sqrt{x_0^2/y_0^2-1},1)$. By tuning the
transversal magnetic field $B_x$ and the current $I$, $y_0$ and thus
the location of the minima can be modified.
 Around these minima, the
potential is parabolic with frequency
\begin{equation}
\omega=\left[
\frac{\mu^2\chi}{m \hbar} \left(\frac{\mu_0I}{2\pi}\right)^2
\frac{1}{y_0^2}\left( \frac{1}{y_0^2}-\frac{1}{x_0^2}\right)
\right]^{1/3} \, .
\end{equation}
Similar to the considerations above, we obtain the potential in units of
$\hbar \omega$,  which depends only on
$d_x=x_0/l_0$, $d_y=y_0/l_0$ and $\chi$,
\begin{eqnarray} \label{dwpot}
\chi V & = &  \left( 1+\frac{\chi d_y^4}{1-d_y^2/d_x^2}
\left\{
\left[
\frac{-y}{(x+d_x)^2+y^2} \right.\right.\right. \nonumber \\
& & \left. \left.\left.
+ \frac{-y}{(x-d_x)^2+y^2} +\frac{1}{d_y}
\right]^2 \right.\right. \nonumber \\
& & \left. \left. +
\left[
\frac{x+d_x}{(x+d_x)^2+y^2} + \frac{x-d_x}{(x-d_x)^2+y^2}
\right]^2
\right\}
 \right)^{1/2}\, . \nonumber \\
\end{eqnarray}
Figure  \ref{fig.2}b) shows
the corresponding bistable potential for the particular case  of
$\chi=0.067$, $I=200 \mu$A, $y_0=100$ nm and $x_0=200$ nm.
The two minima are clearly discerned.
To see how the frequency in the
single well develops if the current in the second wire is turned on,
we introduce the reference frequency $\omega_0$ in the single-well case  with a
fixed current $I$ and a fixed transverse field $B_x$, such that
$y_0=x_0/2$. Then we obtain the ratio
\begin{equation}
\frac{\omega}{\omega_0}
=\left[
\frac{1}{16} \left(\frac{x_0}{y_0}\right)^4
\left( 1-\frac{y_0^2}{x_0^2}\right)
\right]^{1/3} \, .
\end{equation}
For decreasing $B_x$ and keeping $I$ constant, we find that
$\omega$ decreases as shown in Fig.\  \ref{fig.3} (black solid line
and left scale), while the distance
$y_0$ of the atom cloud increases. In the limit $x_0=y_0$, the
two minima merge and the potential becomes quartic and monostable,
implying that $\omega\rightarrow 0$.
For the above parameter set, we find $\omega_0=2 \pi
\times 291$ kHz.
Since one could obtain the same $\omega_0$ for a larger current
$I$ and a correspondingly larger
distance $x_0$, one gets the same trap frequency for
a fixed ratio of $y_0/x_0$. However, $d_x$ and $d_y$ themselves
would change and since the parabolic frequency $\omega$ is fixed,
 only the non-linear corrections to the  parabolic potential
will be modified.
 This in turn influences the height of the potential
barrier and
the tunneling rate between the two wells. Next we study the influence
of the length scale $x_0$ on these two quantities.

Taking the full potential into account, we estimate the barrier
height and the tunneling rate within a simple single-particle WKB 
approximation. The barrier
height $D$ separating the two stable wells,
\begin{equation}
\frac{D}{\hbar \omega} = \chi^{-1}
\left( 1+ \chi d_y^2 \frac{1-d_y/d_x}{1+d_y/d_x}
\right)^{1/2}-\chi^{-1} \, ,
\end{equation}
is shown as a function of $y_0/x_0$
for two values of $I$
in the inset of Fig.\ \ref{fig.3}.
Note that the barrier height is of the order of a few multiples of
the energy gap in the wells, implying that the potential is in the
deep quantum regime, favoring quantum-mechanical tunneling between the two
wells. The corresponding tunneling rate $\Gamma$  for the
lowest-lying pair of energy eigenstates follows in WKB approximation
as
\begin{equation}\label{gamma1}
\frac{\Gamma}{\omega} = e^{-\int_{x_a}^{x_b} dx \sqrt{2 [
V(x,d_y)-1]}} \, ,
\end{equation}
where $x_{a/b}$ are the (dimensionless) classical turning points in the inverted
potential at energy $E=\hbar \omega$, which is approximately the
ground-state energy of a single well. The integral in
Eq.~(\ref{gamma1}) is calculated along the line connecting the two
minima corresponding to $y=y_0$. Results for $\Gamma$ are shown in
Fig.\  \ref{fig.3}  (red solid lines and right scale) as a function of
$y_0/x_0$ for two different values of the current $I$ and the distance
 $x_0$ yielding the same $\omega_0$. Note that for
the smaller current, $I=200$ $\mu$A, $\Gamma$ assumes large
values already for large frequencies $\omega$. This also implies that
the detrimental effects discussed above are less efficient.
On the other hand, for large currents,
the tunneling regime is entered only for much smaller trap
frequencies. For the above parameters, we find $\omega_0=2 \pi
\times 291$ kHz. For the smaller current, the tunneling regime starts at
frequencies of around $\omega=0.37 \omega_0 =2 \pi
\times  108 $ kHz, corresponding
to a temperature of $T=32$ $\mu$K, while for the larger current, the tunneling
regime is entered at $\omega=0.18 \omega_0 =2 \pi
\times  52 $ kHz corresponding to $T=16$ $\mu$K. 
This behaviour illustrates qualitatively (in the single-particle picture) 
one of the benefits of miniaturization. We believe that the features also 
appear in a more detailed consideration involving the atomic correlations which 
is not pursued here. 

A potential drawback of the double wire configuration could be
the transverse  NT deflection
 due to their mutual magnetic repulsion. For an estimate, note that
the NT displacement field $\phi(z,t)$  obeys
the equation of motion $\rho_L \ddot{\phi}=-Y M_I \phi'''' + \mu_0
I^2/(4 \pi x_0)$.  The static solution under the boundary conditions
$\phi(0,L) = \phi'(0,L)=0$
is  $\phi(z) =\mu_0 [Iz(z-L)]^2/(96 \pi Y M_I  x_0)$.
Using again parameters from Ref.~\cite{tubes},
we find the maximum displacement $\phi(L/2)\approx 0.03$~nm
for $L=10\mu$m.  Hence the mutual magnetic repulsion of the NTs is very weak.
Finally, we note that a potential misalignment of the two NT wires is
no serious impediment for the design. Experimentally available techniques could be combined
which allow on the one hand to move a NT on a substrate by an atomic
force microscope \cite{Henk00}, while on the other hand, the NTs can
be suspended and contacted
after being positioned \cite{Kim02}.

\section{Conclusions}

To conclude, we propose a nanoscale waveguide for ultracold atoms
based on doubly clamped suspended nanotubes. We have analyzed this 
scenario from an atom chip point of view. 
All common sources of imperfection can be made sufficiently
small to enable stable operation of the setup.
Two suspended NTs can be combined to create a bistable potential in
the deep quantum regime.
When compared to conventional atom-chip traps employed in
present experiments, such nanotraps
offer several new and exciting perspectives that hopefully
motivate experimentalists to realize this proposal. More refined 
models to study the interplay between the mechanical motion of the 
NTs and the coherent dynamics of the atom cloud are imaginable and could 
establish a link between the field of nanoelectromechanical systems 
and cold atom physics. 

First, rather large trap frequencies can be achieved while at
the same time using smaller wire
currents.  This becomes possible here because both the
spatial size of the atom cloud and its distance to the current-carrying
wire(s) would be reduced to the nanometer scale, and
because NTs allow typical current 
densities of $10 \mu$A$/$nm$^2$, which should be compared to 
the corresponding densities of $10$ nA$/$nm$^2$ in noble metals. 
For the case of a single-well trap, the resulting trap
frequencies go beyond realized chip traps \cite{Folman02}.
Large trap frequencies at low currents are generally
desirable, since detrimental effects like decoherence,
Majorana spin flips, or atom loss will then be significantly reduced. 
Moreover, the faster dynamics of the atoms could lead to the construction of 
fast ''chip circuits''. 

Second, regarding our proposal of a bistable potential with strong
tunneling, the miniaturization towards the nanoscale represents a
novel opportunity to study coherent and incoherent tunneling of a
macroscopic number of cold atoms. The
proposed bistable nanotrap is characterized by 
considerably reduced tunneling distances, thus allowing for large
tunneling rates at large trap frequencies. 
Note that the energy scale associated with
tunneling is larger than thermal energies for realistic temperatures.
Such a bistable device could then switch between the two stable
states on very short time scales enabling the design of fast 
switches. 
Within our proposal the parameters of the bistable potential can be
 tuned over a wide range by modifying experimentally 
 accessible quantities like the current or magnetic fields. 

A third advantage of this proposal results from the 
homogeneity of the currents flowing through the NTs.
As NTs are characterized by long mean free paths, they
often constitute (quasi-)ballistic conductors, where extremely
large yet homogeneous current densities are possible 
which avoids the fragmentation problem \cite{Folman02}.  

Detection certainly constitutes an experimental challenge in this
truly 1D limit.  However, we note that
single-atom detection schemes are currently being developed,
which would also allow to probe the tight 1D cloud here,
e.g., by combining cavity quantum electrodynamics
with chip technology \cite{Reichel02}, or by using additional perpendicular
wires/tubes `partitioning' the atom cloud \cite{reichel}.
This may then allow to study interesting many-body physics in
1D in an unprecedented manner.

\section{Acknowledgments}

We thank A.\ G\"orlitz, Y.\ Kobayashi, C.\ Mora, H.\ Postma, and J.\ Schmiedmayer
for fruitful discussions. V.\ P.\ and M.\ T.\ would like to
thank H.\ Takayanagi and K.\ Semba for the kind hospitality at the NTT
Basic Research Laboratories, where parts of this work have been
accomplished. We acknowledge support by the 
DFG-SFB TR-12 and by the Japanese CREST/JST.

\begin{table}
\caption{Trap frequencies $\omega$, distances $y_0$ of the atomic
cloud from the NT wire, and oscillator lengths $l_0$ for 
$\chi=0.067$ and various $I,d$.
\label{tab.freq}}
\begin{tabular}{|r|r|r|r|r|}
\hline
$I({\rm \mu A})$&$d$&$\omega({\rm kHz})$&$y_0({\rm nm})$&$l_0( {\rm nm})$\\
\hline
1000&10&2$\pi\times$460&144&14\\
250&5&2$\pi\times$460&72&14\\
250&10&2$\pi\times$28.7&576&58\\
100&5&2$\pi\times$73.8&180&36\\
100&10&2$\pi\times$4.6&1440&144\\
50&5&2$\pi\times$18.4&360&72\\
25&5&2$\pi\times$4.6&720&144\\
\hline
\end{tabular}
\end{table}

\begin{table}
\caption{Typical results for the longitudinal size $\ell$ of the
$^{87}$Rb cloud for realistic values of the transversal trap frequency
$\omega$ and  the atom number $N$, where $\omega_z=2 \pi \times 0.1$ kHz.
For $\eta$, see text.
\label{tab.number}}
\begin{tabular}{|r|r|r|r|r|}
\hline
$\omega({\rm kHz})$&$a_{1D}( {\rm
nm})$&N&$\eta$&$\ell( \mu{\rm m})$\\
\hline
2$\pi\times$460&-26.65&30&0.11&7.7\\
2$\pi\times$460&-26.65&50&0.15&10\\
2$\pi\times$73.8&-223&30&0.67&7.3\\
2$\pi\times$73.8&-223&50&0.94&8.7\\
2$\pi\times$73.8&-223&100&1.49&11\\
2$\pi\times$28.76&-603&30&2.55&5.3\\
2$\pi\times$28.76&-603&50&3.58&6.3\\
2$\pi\times$28.76&-603&100&5.72&7.9\\
\hline
\end{tabular}
\end{table}

\end{document}